# Magneto-impedance of glass-coated Fe-Ni-Cu microwires

J. Wiggins, H. Srikanth[a], K.-Y. Wang, L. Spinu and J. Tang
*Advanced Materials Research Institute, University of New Orleans, New Orleans, LA 70148*

The magneto-impedance (MI) of glass-coated Fe-Ni-Cu microwires was investigated for longitudinal radio-frequency (RF) currents up to a frequency of 200 MHz using an RF lock-in amplifier method. The MI, defined as $\Delta Z/Z = [Z(H)-Z(H=0.3T)]/Z(H=0.3T)$, displays a peak structure (negative MI) at zero field for RF currents with frequencies less than 20MHz and this crosses over to a sharp dip (positive MI) at higher frequencies. This crossover behavior is ascribed to the skin-depth-limited response primarily governed by the field-dependence of the permeability. Large saturation fields (300 to 600 Oe) and other anomalies indicate the possible influence of giant magneto-resistance (GMR) on the MI.

## I. INTRODUCTION

The magneto-impedance (MI) of thin soft ferromagnetic wires has been studied extensively over the past few years and is a topic of great current interest[1-3]. Large changes in MI often referred to as giant magneto-impedance (GMI) have been observed in a wide range of materials primarily in the forms of amorphous or nanocrystalline wires, ribbons and films. GMI holds a lot of promise in technologically important applications like field sensors and magnetic recording heads. Systematic studies of MI are also vital as they essentially determine the response of materials and consequently, devices, operating at RF and microwave frequencies.

The MI effect consists of a significant change in the impedance of a soft magnetic conductor, driven by a high frequency current, when it is placed in a static magnetic field. In the case of a cylindrical wire, a transverse field geometry is normally employed with the static field ($H$) along the axial direction and the RF current passing through the wire also in the same direction, thus setting up an oscillatory RF field ($H_{rf}$) around the circumference of the wire. When $H \leq H_K$, where $H_K$ is the circumferential anisotropy field, the MI effect itself can be considered a purely classical phenomenon resulting from the interaction between $H_{rf}$ and the magnetic domain structure in the sample. For $H > H_K$, other phenomena such as ferromagnetic resonance (FMR), may drive the MI effect.

The complex impedance of a cylindrical magnetic conductor can be expressed as[4]:

$$Z = R_{dc} ka J_0(ka)/2 J_1(ka), \quad (1)$$

where $R_{dc}$ is the dc resistance of the wire, $a$ its radius, $J_0$ and $J_1$ are Bessel functions of the first kind, and $k$ is the radial propagation constant which is related to the effective skin depth ($\delta$) through

$$k = (1-j)/\delta. \quad (2)$$

The skin depth, in turn, is related to the material resistivity ($\rho$), permeability ($\mu$) and frequency ($\omega$) of the RF current, and can be written as:

$$\delta = (2\rho/\mu\omega)^{1/2}. \quad (3)$$

From (1)-(3), it can be seen that $Z(H)$ is directly governed by the change in permeability, $\mu(H)$ and resistivity, $\rho(H)$. Generally in soft ferromagnetic wires, the magnetoresistance, $\Delta\rho(H)/\rho(H=0)$, is small and the MI effect is almost entirely dominated by $\mu(H)$.

The MI effect has been investigated in a number of Fe- and Co- based wires and changes ranging from a few % to several 100% have been reported, with the largest MI seen in amorphous Co-based wires with nearly zero magnetostriction[5].

In this paper, we report MI measurements on glass-coated Fe-Ni-Cu microwires using an RF lock-in amplifier technique. The MI itself is a figure of merit that can be defined in a number of ways. We have defined it as

$$\Delta Z/Z = [Z(H)-Z(H_{max})]/Z(H_{max}), \quad (4)$$

where $H_{max}$ is the maximum field applied. In our case, $H_{max} = 0.3\ T$.

These samples differ from the majority of soft ferromagnetic wires studied in the sense that they are more granular in nature due to the low solubility of Ni and Fe in Cu while the alloy is formed. Recently, Wang et al.[6] reported observation of giant magnetoresistance (GMR) in these wires which is quite interesting as soft ferromagnetic materials do not generally exhibit GMR. Our MI measurements were motivated by the possibility of studying the phenomenon in a system where the MR also shows large changes with magnetic field.

## II. EXPERIMENTAL

---
[a] Corresponding author; Electronic mail: sharihar@uno.edu

The Fe-Ni-Cu alloy (Fe 20%, Ni 20% and Cu 60%) was prepared by induction melting in a Pyrex glass tube. Glass-coated microwires of diameter ~5μm were drawn from this melt using Taylor's technique. Structural characterization was done using X-ray diffraction and transport measurements were performed in a commercial Physical Property Measurement System (Quantum Design). Negative magnetoresistance (characteristic of GMR) of about 8% was measured on as-cast microwires with no trace of saturation up to 9 Tesla. Details of sample preparation and MR measurements have been presented elsewhere[6]. The M-H loop measured with a SQUID magnetometer is shown in Fig. 1 indicating that the coercivity is small, which is a characteristic of soft magnetic materials. The shape of the curve indicates that the approach to saturation is very broad as the magnetic field is increased.

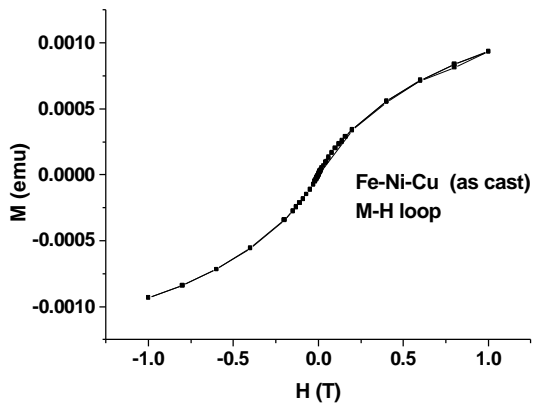

Fig. 1. Magnetization curve for as-cast 5 mm thick Fe-Ni-Cu microwire obtained using a SQUID magnetometer (Quantum Design)..

The MI measurements were done using a customized RF experimental set-up that we developed for complex impedance studies at cryogenic temperatures[7]. Samples were mounted on a fixture located on a low temperature insert that can be placed in the Physical Property Measurement System (PPMS). This provides the platform for varying the temperature and static magnetic field (*H*). A four-probe geometry was employed to pass RF current along the axial direction of the wire and measure the voltage. A RF current with an RMS amplitude of *10mA* was provided by the built-in oscillator of a Model SR844 lock-in amplifier (Stanford Research). The current was verified to be constant over the frequency range of measurement up to 200MHz. The lock-in also measured the in-phase and quadrature components of the voltage drop across the microwire.

### III. RESULTS AND DISCUSSION

Knowing the amplitude of the driving current, the change in amplitude of the impedance can be obtained. This is plotted as $DZ/Z$ as defined in equation (4) with $H_{max} = 0.3T$. Characteristic data sets with this quantity plotted against the static field (H) were obtained for several fixed frequency RF currents. All the data reported here were taken at a temperature of 100K. We did measure similar data sets at two other temperatures (200K and 50K). Although the zero-field impedance values changed consistently with the temperature-dependent resistivity of the sample, the field-dependence of $DZ/Z$ did not show any significant variation from the data presented here. Fig. 2 shows the MI data for the wire at three different frequencies (10, 20 and 50 *MHz*). The striking features clearly seen in the data are the presence of sharp changes around $H = 0$ and the crossover from a peak to a dip structure at higher frequencies. The dashed vertical lines in Fig. 2 indicate the static field region over which these sharp changes occur and beyond which saturation is attained. Typically this field is between ± 200 to 300 Oe.

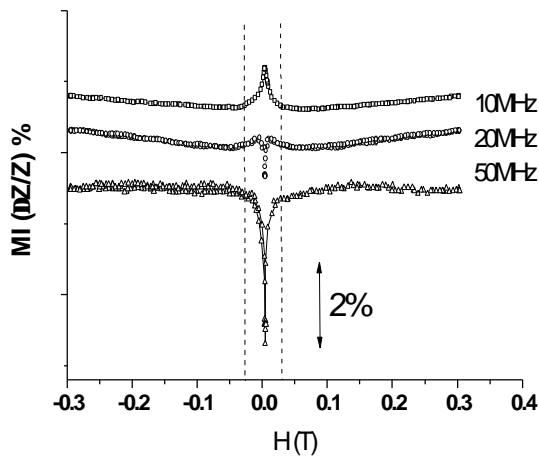

Fig. 2. Magneto-impedance of the wire at three different frequencies. Notice the crossover in shape near zero field.

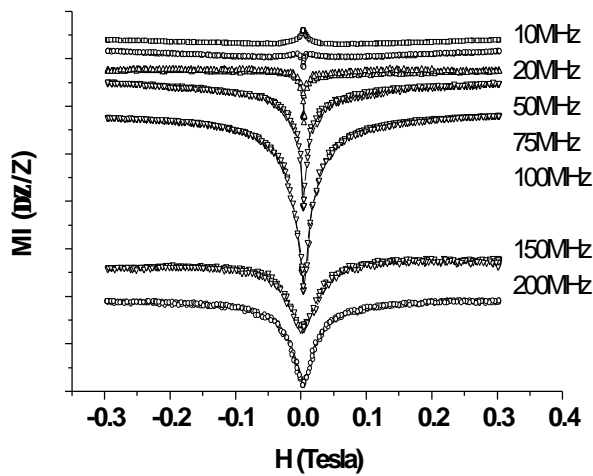

Fig. 3. MI at several frequencies up to 200MHz. The largest MI change is about 10% for f = 100MHz. Data are normalized with respect to saturation fields and relatively shifted for clarity.

In Fig. 3, the MI data at higher frequencies are also presented. The data are normalized and relatively shifted for clarity. As can be seen, the zero-field dip becomes more pronounced but at the same time it also broadens leading to an apparent increase in the saturation fields up to ± 600 Oe. The largest positive MI percentage of about 10% was obtained for an RF current at 100MHz. At the highest frequencies measured, the percentage change in MI actually reduces. Analyzing the MI effect is complicated as it is influenced by a number of factors. While the dominant contribution is due to the rapid change in circumferential permeability, many other parameters like microstructure, anisotropy, domain configuration, induced stress in the material etc. contribute to the overall response[4]. Systematic experiments are needed to separately evaluate the individual contributions from each of these parameters. Here, we attempt to present a simple description of some of the key features observed in our MI data.

We now focus on the shape changing over from a peak to a dip at higher frequencies, as plotted in Fig. 2. In glass-coated microwires, the anisotropy is due to magnetostatic and magnetoelastic effects. The former results from the strong shape anisotropy in a thin cylindrical wire and latter arises from large residual stresses at the surface due to the glass coating. A core-shell model has been effective in the treatment of MI in these types of wires[8-10]. The region of the wire sampled by the RF current decreases as the frequency increases and this is related by the well-known expression for skin-depth, given in equation (3). Depending on the material resistivity and permeability, it is possible to have a situation where the RF current flows in the core of the wire at low frequencies and only in a thin shell close to the surface at high frequencies. The crossover in shape in the data presented in Fig. 2 can be ascribed to this situation where the influence due to the two different anisotropies govern the MI effect.

A simple estimation of the skin depth from equation (3) validates the plausibility of this scenario. Taking the measured dc resistivity of the sample to be 25μΩ-cm at 100K and assuming a relative permeability between 500 and 1000 (which are reasonable limits for this alloy), the skin depth is estimated between 2.5 to 3.5 μm for a frequency of 10MHz and reduces to less than 1 μm for frequencies above 100MHz. These values are well within the cross-section of the wire around 5 μm.

In general, two types of magnetization processes contribute to the MI. These are domain wall displacement and domain magnetization rotation[4]. At our measurement frequencies, the main contribution is from the latter process as the domain wall relaxation frequencies are likely to be much lower. The influence on MI due to the two different domain structures can be cast in terms of a phenomenological model based on magnetization rotation. The crossover from the MI peak to dip at $H=0$ is reproduced considering different angles (0 to 90°) that the anisotropy axis in the two types of domains makes with the transverse RF field[4,11].

While the crossover with frequency can be qualitatively explained, there are some aspects of our MI data that differ considerably from other results on soft ferromagnetic wires. These are: (a) the MI saturation appears to take place at much larger fields (typically 300 to 600 Oe) (see Fig. 3). (b) The overall change in MI is not very large and typically less than 10%. (c) The trend of increase in MI % up to 100MHz followed by a reduction at higher frequencies as seen in Fig. 3.

An important aspect that we believe plays a role here, is the influence of GMR. MR of the sample is usually ignored as it is much smaller compared to the change in permeability. However, the large GMR noted in these wires[6] due to their granular nature suggests that the MI is likely to be affected by it. Since the impedance $Z \propto (\mu\rho)^{1/2}$, changes in both μ and ρ should be considered. Saturation of MR takes place at a much larger field compared to the permeability saturation. So a combination of both these effects could lead to broader MI variation as seen in our data. Systematic measurements on annealed samples are needed to explore these issues further. Impedance matching problems associated with this lock-in amplifier method make it difficult to obtain the MI dependence as a function of smoothly varying frequency. Using an impedance analyzer helps eliminate signal reflections in the cables and make broadband frequency sweeps. We are currently extending our measurement capabilities for MI studies using a Hewlett-Packard HP4291B impedance analyzer.

## ACKNOWLEDGEMENTS

This work is supported by DARPA through grant No. MDA 972-97-1-0003. The authors would like to thank the Director of AMRI, Dr. C. J. O'Connor for his interest and support.